\documentclass[trackchanges]{aastex701}
\usepackage{comment}
\usepackage{hyperref}
\hypersetup{colorlinks=true,linkcolor=blue,citecolor=blue}

\shorttitle{FAST detection of PSR J2238+5903}
\shortauthors{Zhang et al.}

\begin{document}

\title{FAST Discovery of $\mu$Jy Radio Pulsations from PSR J2238+5903, Providing a DM Distance Anchor for the Candidate TeV Halo 1LHAASO J2238+5900}

\correspondingauthor{ Hui Zhu, Guanhong Lin, Songzhan Chen}
\email{ zhuhui@nao.cas.cn, lingh@bao.ac.cn, chensz@ihep.ac.cn}

\author[orcid=0000-0002-2940-4821]{Jianli Zhang}
\affiliation{National Astronomical Observatories, Chinese Academy of Sciences, 100101 Beijing, China; }
\affiliation{University of Chinese Academy of Sciences, Beijing 100049, People’s Republic of China}
\email{jlzhang@bao.ac.cn}

\author{Hui Zhu}
\affiliation{National Astronomical Observatories, Chinese Academy of Sciences, 100101 Beijing, China; }
\affiliation{University of Chinese Academy of Sciences, Beijing 100049, People’s Republic of China}
\email{jlzhang@bao.ac.cn}

\author[orcid=0009-0006-0364-4161]{Guanhong Lin}
\affiliation{National Astronomical Observatories, Chinese Academy of Sciences, 100101 Beijing, China; }
\affiliation{University of Chinese Academy of Sciences, Beijing 100049, People’s Republic of China}
\email{jlzhang@bao.ac.cn}

\author{Dejia Zhou}
\affiliation{National Astronomical Observatories, Chinese Academy of Sciences, 100101 Beijing, China; }
\affiliation{University of Chinese Academy of Sciences, Beijing 100049, People’s Republic of China}
\email{jlzhang@bao.ac.cn}

\author{Yuting Chu}
\affiliation{National Astronomical Observatories, Chinese Academy of Sciences, 100101 Beijing, China; }
\affiliation{University of Chinese Academy of Sciences, Beijing 100049, People’s Republic of China}
\email{jlzhang@bao.ac.cn}

\author{Songzhan Chen}
\affiliation{Key Laboratory of Particle Astrophysics, Institute of High Energy Physics, Chinese Academy of Sciences, 100049 Beijing, People's Republic of China; }
\affiliation{Tianfu Cosmic Ray Research Center, 610000 Chengdu, Sichuan, People's Republic of China}
\affiliation{University of Chinese Academy of Sciences, 100049 Beijing, People's Republic of China}
\email{chensz@ihep.ac.cn}

\author{Min Zha}
\affiliation{Key Laboratory of Particle Astrophysics, Institute of High Energy Physics, Chinese Academy of Sciences, 100049 Beijing, People's Republic of China; }
\affiliation{Tianfu Cosmic Ray Research Center, 610000 Chengdu, Sichuan, People's Republic of China}
\affiliation{University of Chinese Academy of Sciences, 100049 Beijing, People's Republic of China}
\email{zham@ihep.ac.cn}

\author{WenJun Huang}
\affiliation{School of Physics and Astronomy, Sun Yat-sen University, Zhuhai 519082, People's Republic of China}
\affiliation{CSST Science Center for the Guangdong-Hong Kong-Macao Greater Bay Area, Zhuhai 519082, People's Republic of China}
\email{jlzhang@bao.ac.cn}

\author{ZiWei Ou}
\affiliation{Tsung-Dao Lee Institute, Shanghai Jiao Tong University, Shanghai 201210, China}
\email{jlzhang@bao.ac.cn}

\author{P. H. Thomas Tam}
\affiliation{School of Physics and Astronomy, Sun Yat-sen University, Zhuhai 519082, People's Republic of China}
\affiliation{CSST Science Center for the Guangdong-Hong Kong-Macao Greater Bay Area, Zhuhai 519082, People's Republic of China}
\email{jlzhang@bao.ac.cn}

\author{Sha Wu}
\affiliation{Key Laboratory of Particle Astrophysics, Institute of High Energy Physics, Chinese Academy of Sciences, 100049 Beijing, People's Republic of China; }
\affiliation{Tianfu Cosmic Ray Research Center, 610000 Chengdu, Sichuan, People's Republic of China}
\email{jlzhang@bao.ac.cn}

\author{Qiang Yuan}
\affiliation{Purple Mountain Observatory, Chinese Academy of Sciences, Nanjing 210023, People's Republic of China}
\affiliation{Key Laboratory of Dark Matter and Space Astronomy, Chinese Academy of Sciences, Nanjing 210023, People's Republic of China}
\email{jlzhang@bao.ac.cn}

\author{Yi Zhang}
\affiliation{Purple Mountain Observatory, Chinese Academy of Sciences, Nanjing 210023, People's Republic of China}
\affiliation{Key Laboratory of Dark Matter and Space Astronomy, Chinese Academy of Sciences, Nanjing 210023, People's Republic of China}
\email{jlzhang@bao.ac.cn}


\begin{abstract}
We report the first detection of radio pulsations from PSR J2238+5903, a gamma-ray pulsar spatially coincident with the extended TeV source 1LHAASO J2238+5900. Our 3000 s FAST L-band observation reveals a weak periodic signal at the known Fermi-LAT spin period, with $P=162.76568$ ms and $\mathrm{DM}=247.5\pm3.0~\mathrm{pc~cm^{-3}}$. The signal is independently confirmed by both FFT-based and Fast Folding Algorithm searches. The radiometer equation gives a flux density of $S_{1250}\simeq3\,\mu$Jy, placing PSR J2238+5903 among the faintest radio-detected Fermi pulsars. Interpreting the DM with Galactic electron-density models gives $d_{\rm DM}=7.4\pm3.9$ kpc. At this distance, the LHAASO WCDA 39\% containment radius corresponds to a characteristic diameter of $\sim132$ pc, and the $>1$ TeV luminosity is $L_{\rm TeV}\simeq7.1\times10^{34}$ erg s$^{-1}$, about 8\% of the pulsar's spin-down power. The radio DM thus provides the first pulsar-specific distance constraint for assessing whether 1LHAASO J2238+5900 is a young relic-PWN / TeV-halo transition system.
\end{abstract}


\section{Introduction}

TeV halos, discovered by the HAWC Observatory around the middle-aged pulsars Geminga and Monogem \citep{Abeysekara2017}, represent a new class of Galactic very-high-energy (VHE, $E>0.1$ TeV) gamma-ray sources. These extended structures are thought to arise from inverse Compton scattering of interstellar radiation fields by relativistic electrons and positrons that have escaped from the pulsar wind nebula (PWN) and diffused into the surrounding medium \citep{Linden2017}. The inferred diffusion coefficients are orders of magnitude smaller than the Galactic average, suggesting that cosmic-ray transport is inhomogeneous in the vicinity of pulsars. Recent stacking analyses of middle-aged pulsars suggest that TeV halos may be a common phenomenon \citep{HAWC_stacking2025}, yet the sample of well-characterized individual halos remains limited, and their diversity is not fully understood. For example, LHAASO detected a point-like ultra-high-energy source, 1LHAASO J1740+0948u, near PSR J1740+1000, with a significant spatial offset from the pulsar and a small extension upper limit that disfavors a classical TeV halo interpretation \citep{LHAASO_J1740_Innovation}.

The Large High Altitude Air Shower Observatory (LHAASO) has identified numerous extended VHE sources in its first catalog \citep{Cao2024}, some of which have been interpreted as pulsar halos or evolved PWNe associated with middle-aged pulsars, including 1LHAASO J0622+3754/PSR J0622+3749 and 1LHAASO J0249+6022/PSR J0248+6021 \citep{LHAASO_J0621+3749, LHAASO_J0248+6021}. Among them, 1LHAASO J2238+5900 is flagged as likely associated with a pulsar, with a chance coincidence probability of $\sim$0.03\% \citep{Cao2024}, and identified by \citet{Zheng2024} as a candidate TeV halo based on the lack of significant residual GeV emission and its hard TeV spectrum. The TeV emission is notably extended ($\sim$0.5$^\circ$ scale) and detected by both WCDA (1--25 TeV) and KM2A ($>25$ TeV) \citep{Cao2024}. The source is spatially coincident with the Fermi-LAT pulsar PSR J2238+5903 (4FGL J2238.5+5903), which has a spin period $P=162.74$ ms, spin-down luminosity $\dot{E}=8.9\times10^{35}$ erg s$^{-1}$, and characteristic age $\tau_{\rm char}=26.6$ kyr \citep{Smith2023}. Despite its high spin-down power, the pulsar had never been detected in radio, with a deep 1400 MHz flux density upper limit $<0.011$ mJy \citep{Ray2011}, and a 1250 MHz flux density upper limit $<0.005$ mJy recently by FAST \citep{Wangsq2025}, classifying it as radio-quiet ($S_{1400}<30\,\mu$Jy) following the Fermi convention \citep{Smith2023}. Many gamma-ray pulsars remain undetected at radio wavelengths, and whether this reflects intrinsic radio weakness or unfavorable beaming geometry remains debated. For pulsars associated with TeV halos, this typically precludes a dispersion measurement(DM) and thus a reliable distance—a critical missing parameter for characterizing the extended TeV emission.

Alternative distance estimates for PSR J2238+5903 exist in the literature, though none are based on direct measurements. Based on kinematic arguments, \citet{Dincel2024} proposed an association between the pulsar and the young open cluster Berkeley 97, for which independent photometric and Gaia-based studies consistently yield a distance of $\sim$3.0 kpc (e.g., \citealt{Dias2021}). In a complementary approach, \citet{Anguner2025} applied the Fundamental Plane relation---a correlation between gamma-ray luminosity and pulsar intrinsic properties---to Fermi-LAT data and predicted a distance of $2.35$--$3.12$ kpc for this pulsar using various machine learning models. While these two independent indirect methods converge on a similar distance range, both rely on model-dependent assumptions and lack direct observational confirmation. Consequently, the true distance to PSR J2238+5903 has remained uncertain.

The lack of a radio detection has prevented the measurement of the pulsar's DM---a direct observable tied to the pulsar itself---which would provide an independent distance constraint. To address this, we initiated a Five-hundred-meter
Aperture Spherical radio Telescope (FAST; \citep{Nan2011,Jiang2020}) campaign (PI: H. Zhu) in 2021 to search for radio counterparts of LHAASO TeV halo candidates. An initial detection was reported in Astronomer's Telegram \#17756 \citep{ATel17756}. In this Letter, we present the first refereed radio detection of PSR J2238+5903, providing the first direct measurement of its DM and thereby an independent, model-dependent distance anchor for determining the physical properties of 1LHAASO J2238+5900.

\section{Observations and Data Analysis}

\subsection{FAST Observations}

We observed PSR J2238+5903 on 2021 August 14 using the FAST L-band 19-beam receiver (1.05--1.45 GHz) in tracking mode. The central beam was pointed at 4FGL J2238.5+5903 of the Fermi-LAT position (RA $=22^{\rm h}38^{\rm m}28.15^{\rm s}$, Dec $=+59^{\circ}03'44.8''$, J2000). The total on-source integration time was 3000 s. Data were recorded in search-mode PSRFITS format with 8-bit sampling, four polarizations, 4096 frequency channels, and a sampling interval of 49.152 $\mu$s \citep{Jiang2020, Hotan2004}.

\subsection{Periodicity Search}

We performed periodicity searches for PSR J2238+5903 using both FFT-based and FFA-based approaches. FFT-based periodic pulsation searches were performed with the PRESTO pulsar search suite \citep{Ransom2011}. Radio frequency interference (RFI) was first identified and excised using the PRESTO routine \texttt{rfifind}. A detailed characterization of the RFI environment at FAST is provided in \citet{Jiang2020}.

The script \texttt{DDplan.py} was employed to determine the optimal dedispersion strategy. The data were dedispersed over a DM range of 0 to 3000 pc cm$^{-3}$ with a step size that increased with DM to optimize computational efficiency: 0.1, 0.2, 0.3, 0.5, 1.0, 3.0, 5.0~${\rm pc~cm^{-3}}$ for DM ranges starting at 0, 37, 113.4, 189, 322.2, 544.2 and 1432.2~${\rm pc~cm^{-3}}$, respectively.

The dedispersed time series were subsequently Fourier transformed for FFT-based periodicity searches. We used the PRESTO routine \texttt{accelsearch} to search for periodic signals, summing up to $h=16$ harmonics and allowing a maximum drift parameter of $z_{\mathrm{max}}=0$ (since PSR J2238+5903 is isolated). Candidate signals were then sifted using the PRESTO script \texttt{ACCEL\_SIFT.py} with a significance threshold of sigma $>4.0$ to consolidate multiple detections of the same candidate across different DMs. Promising candidates were subsequently folded using PRESTO's \texttt{prepfold} command, and the resulting diagnostic plots were examined visually.

To further verify the robustness of the detection, we performed an independent blind search using the Fast Folding Algorithm (FFA) \citep{FFA}, implemented in the RIPTIDE package. The FFA search was applied to the dedispersed time series, searching trial periods from 0.05 to 0.4 s with a signal-to-noise ratio threshold of 7.0. The FFA search recovered the same period and DM as the FFT-based search, independently confirming the astrophysical nature of the signal.

\subsection{Single-Pulse Search}

Rotating radio transients and some pulsars emit sporadic single pulses that may evade detection in periodicity searches \citep{McLaughlin2006}. We therefore conducted a single-pulse search using the GPU-accelerated pipeline ASTROFLOW \citep{Lin2026}, which employs a deep learning detector trained on single-pulse data. The search covered a DM range of 0 to 3000 pc cm$^{-3}$ with a detection threshold of S/N $>6$. No significant single pulses were detected, consistent with the pulsar's weak average flux density.

\section{Results}

\subsection{Radio Detection and Distance}

\begin{figure}[htb]
\centering
\begin{minipage}{0.9\columnwidth}
    \centering
    \includegraphics[width=\textwidth]{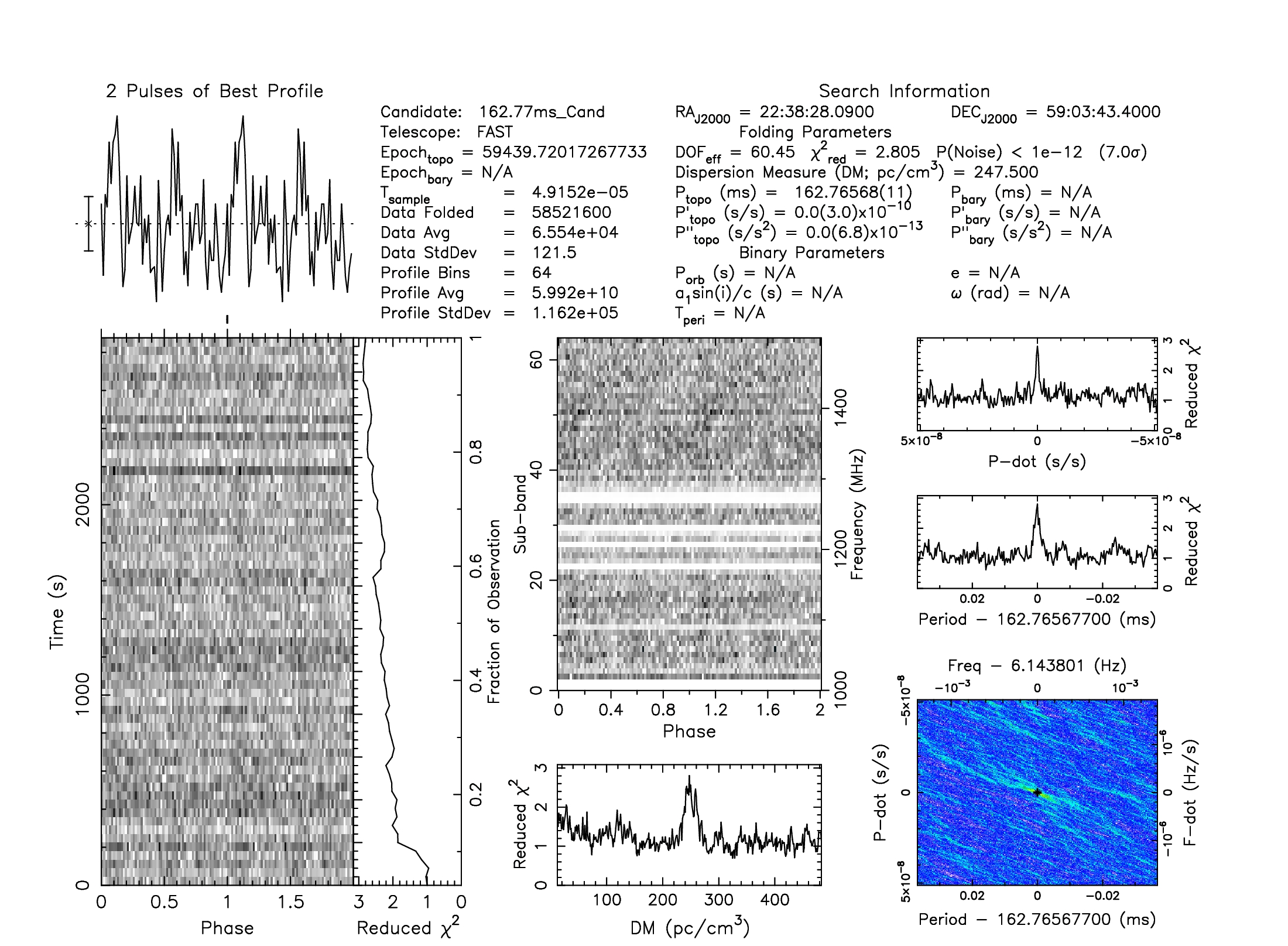}
    \vspace{2mm}
    (a) PRESTO (FFT)
\end{minipage}

\vspace{4mm}

\begin{minipage}{0.9\columnwidth}
    \centering
    \includegraphics[width=\textwidth]{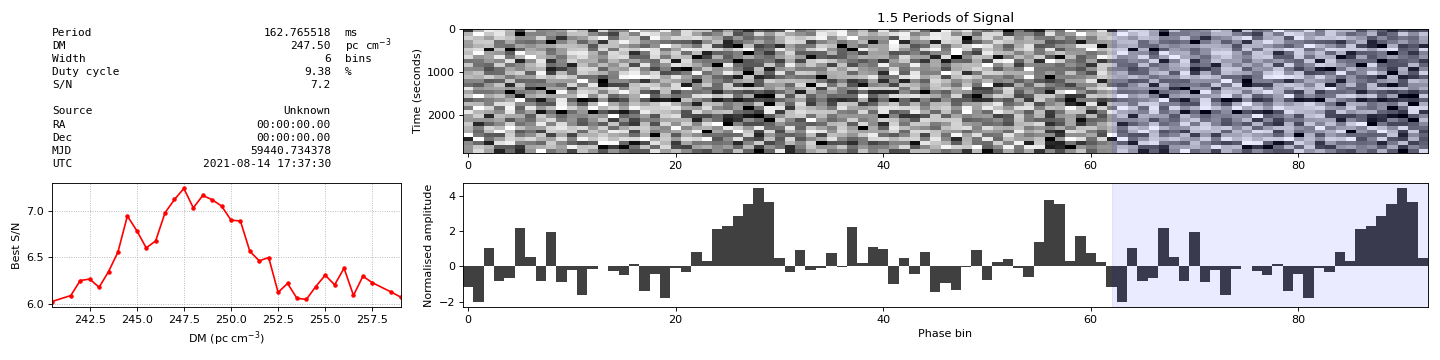}
    \vspace{2mm}
    (b) RIPTIDE (FFA)
\end{minipage}

\caption{FAST detection of PSR J2238+5903 confirmed by two independent search methods. 
Panel (a): PRESTO detection via FFT-based periodicity search, showing the folded pulse 
profile at \(P=162.76568(11)\) ms and DM \(=247.5\) pc cm\(^{-3}\) with S/N \(\approx7.0\). 
Panel (b): independent confirmation by the RIPTIDE Fast Folding Algorithm, recovering 
the same period and DM with S/N \(\approx7.2\). The combination of two independent search 
methods, together with the detection of a half-period harmonic (S/N\(\approx8.2\)), strongly supports the astrophysical origin of the pulsation.}
\label{fig:detection}
\end{figure}

PSR J2238+5903 has two distinct components in its gamma-ray profile \citep{Smith2023}. We conducted a 3000 s FAST observation on 2021 August 14. The signal was detected with S/N $\approx7.0$ at the optimal DM of $247.5\pm3.0$ pc cm$^{-3}$. Figure~\ref{fig:detection} shows the detection from both search methods. Panel (a) presents the folded profile from the PRESTO FFT-based search at the candidate period and DM; the nonzero-DM peak serves as the primary discriminator against zero-DM RFI. The folded pulse profile yielded a spin period $P = 162.76568(11)$ ms at epoch MJD 59440(the day of FAST observation), agrees with the gamma-ray timing solution propagated to the FAST epoch to within $\approx10^{-5}$ in fractional period. The detection was independently recovered by an FFA blind search applied to the same FAST data, with S/N $\approx7.2$ at the same DM (Figure~\ref{fig:detection}, panel b). A signal near half the trial period (S/N $\approx8.2$) was also detected, consistent with the harmonic structure expected from the folded pulse morphology.

As a further check against RFI, we searched the other 18 beams of the FAST 19-beam receiver that were recorded simultaneously. No signal was found at the candidate period or DM in any off-source beam, confirming that the detection is spatially localized to the target position and is not of terrestrial origin.

The minimum detectable flux density for our observation can be estimated using the radiometer equation \citep{Lorimer2005}:

\begin{equation}
S_{\rm min} = \beta \frac{(S/N)_{\rm min} T_{\rm sys}}{G\sqrt{n_{\rm pol} t_{\rm obs} \Delta f}} \sqrt{\frac{W}{P-W}},
\label{eq:radiometer}
\end{equation}

with $\beta=1$ (8-bit sampling), $T_{\rm sys}\approx27$ K, $G=16$ K Jy$^{-1}$ \citep{Jiang2020}, $n_{\rm pol}=2$, $\Delta f=350$ MHz (after RFI excision), $t_{\rm obs}=3000$ s, duty cycle $W/P=0.1$, and $(S/N)_{\rm min}=7.0$. This gives a flux density of $S_{1250} \approx 2.7\ \mu$Jy. Adopting a conservative uncertainty, we report $S_{1250}=3\,\mu$Jy.

The measured DM gives model-dependent distances of $d_{\mathrm{NE2001}} \sim 9.0$~kpc \citep{Cordes2002}, $d_{\mathrm{YMW16}} \sim 7.0$~kpc \citep{Yao2017}, and $d_{\mathrm{NE2025}} \sim 6.3$~kpc \citep{NE2025}. These are not geometric distances; they inherit the systematic uncertainties of the electron-density models along this low-latitude sightline. The mean value of the three estimates is $d_{\mathrm{mean}} = 7.43$~kpc, and their standard deviation is $\sigma_{\mathrm{sys}} = 1.17$~kpc. Following the conservative uncertainty prescription adopted in \citet{Deller2019} and \citet{Zhanglei2026}, we assign a 50\% model uncertainty, $\sigma_{\mathrm{stat}} = 0.50 \times d_{\mathrm{mean}} = 3.72$~kpc. Adding these in quadrature gives $\sigma_{\mathrm{tot}} = \sqrt{\sigma_{\mathrm{sys}}^2 + \sigma_{\mathrm{stat}}^2} = 3.90$~kpc. We therefore adopt $d_{\mathrm{DM}} = 7.4 \pm 3.9$~kpc.

\subsection{Physical Properties Derived from the DM Distance}

The measured DM provides the first direct, pulsar-tied distance constraint for PSR J2238+5903. Adopting $d=7.4$ kpc as a fiducial scaling distance, the 4FGL energy flux $G_{100}=(6.6\pm0.24)\times10^{-11}$ erg cm$^{-2}$ s$^{-1}$ \citep{Smith2023} gives

\begin{equation}
\frac{L_\gamma}{\dot{E}} \simeq
0.49\, f_\Omega \left(\frac{d}{7.4~{\rm kpc}}\right)^2,
\label{eq:gev_efficiency}
\end{equation}

where we have assumed $f_\Omega=1$. This GeV efficiency is within the broad range ($\sim$0.1--100\%) observed for young rotation-powered pulsars \citep{Smith2023}. For the TeV band, integrating the WCDA power-law spectrum ($dN/dE=N_0(E/3\,{\rm TeV})^{-\Gamma}$ with $N_0=1.91\times10^{-13}$ cm$^{-2}$ s$^{-1}$ TeV$^{-1}$ and $\Gamma=2.39$; \citealt{Cao2024}) above 1 TeV gives

\begin{equation}
\frac{L_{>1{\rm TeV}}}{\dot{E}} \simeq
0.080 \left(\frac{d}{7.4~{\rm kpc}}\right)^2.
\label{eq:tev_efficiency}
\end{equation}

This TeV efficiency is comparable to those inferred for other TeV halos such as Geminga ($\sim$1\%) and Monogem ($\sim$2\%) \citep{Abeysekara2017}, and is consistent with the expectation that a significant fraction of the pulsar's spin-down power is channeled into the extended TeV emission. The WCDA 39\% containment radius, $r_{39}=0.51^\circ$, corresponds to a characteristic diameter

\begin{equation}
D_{39}=2d\tan r_{39}
\simeq 132 \left(\frac{d}{7.4~{\rm kpc}}\right)~{\rm pc}.
\label{eq:diameter}
\end{equation}

This characteristic diameter of $\sim132$ pc is comparable to that of the giant gamma-ray nebula HESS J1825--137 ($\sim100$ pc; \citealt{HESS2018}). However, $r_{39}$ is a containment scale rather than the full extent of the source, and the KM2A scale at higher energies ($r_{39}=0.43^\circ$) is somewhat smaller, qualitatively consistent with energy-dependent cooling in a diffusive particle population. For comparison, adopting the previous pseudo-distance of 2.83 kpc would yield significantly smaller values: $L_{\rm TeV}\simeq1.0\times10^{34}$ erg s$^{-1}$, $L_{\rm TeV}/\dot{E}\simeq1.2\%$, and $D_{39}\simeq50$ pc, underscoring the strong dependence of the inferred physical properties on the adopted distance.

\section{Discussion}

\subsection{X-ray Counterpart}

An X-ray counterpart to PSR J2238+5903 has been identified by \citet{Shibanov2025} using eROSITA and Chandra, with a soft spectrum ($\Gamma=4.0\pm0.4$). Their modeling adopted a gamma-ray pseudo-distance of $d=1.9$ kpc; a self-consistent reanalysis with an $N_{\rm H}$ prior appropriate for our DM distance $d=7.4$ kpc is needed to obtain firm constraints on the thermal emission. Deeper X-ray observations are also required to search for extended PWN emission that might accompany the TeV halo.

The TeV halo interpretation of 1LHAASO J2238+5900 is independently supported by \citet{Zheng2024}, who found no significant residual GeV emission in the off-pulse Fermi-LAT data, ruling out contamination from other Galactic sources. The large angular extent and the absence of an associated supernova remnant or bright PWN are also consistent with this interpretation.

\subsection{A Back-of-the-Envelope Diffusion Estimate}

Following the method used for the LHAASO halo J0621+3755 \citep{Zhang2021}, we estimate the diffusion coefficient using the physical scale derived from the KM2A extension, as the cooling time is evaluated at the corresponding photon energy. The LHAASO catalog reports a KM2A reference energy of $E_0=50$ TeV for 1LHAASO J2238+5900 \citep{Cao2024}. Adopting the approximate scaling between photon and electron energies used in \citet{Zhang2021}, this corresponds to electrons with energy $\sim160$ TeV, for which the synchrotron cooling time is $\tau_{\rm cool}\approx5.5$ kyr. Under the simplified assumption of isotropic diffusion, $R=\sqrt{2D\tau_{\rm cool}}$, the KM2A 39\% containment scale $r_{39}=0.43^\circ$ gives a physical radius $R = d\tan(0.43^\circ) \simeq 55.5$ pc at $d=7.4$ kpc, yielding

\begin{equation}
D \approx \frac{R^2}{2\tau_{\rm cool}} \simeq 8.5\times10^{28}\ \text{cm}^2\ \text{s}^{-1}.
\label{eq:diffuse}
\end{equation}

This estimate is of the same order as the suppressed diffusion coefficients inferred for Geminga ($\sim10^{28}$ cm$^2$ s$^{-1}$) and LHAASO J0621+3755 ($\sim10^{28}$ cm$^2$ s$^{-1}$), and remains much smaller than the Galactic average of $\sim10^{30}$--$10^{31}$ cm$^2$ s$^{-1}$ inferred from cosmic-ray secondary-to-primary ratios \citep{Yuan2017}. We emphasize that this is an order-of-magnitude estimate; a full transport model accounting for injection history, magnetic field, and projection effects is beyond the scope of this Letter.

\subsection{Comparison with other Distance Estimates}

Our DM-based fiducial distance of $d_{\rm DM}=7.4\pm3.9$ kpc is larger than previous indirect estimates of $\sim$2.3--3.1 kpc, including the proposed Berkeley 97 association, the Fundamental Plane estimate, and the ATNF pseudo-distance. The difference should not be interpreted as a definitive rejection of those estimates. The DM is a direct observable tied to the pulsar, but its conversion to distance depends on the Galactic free-electron distribution, which is uncertain along low-latitude sightlines and may be affected by unmodeled H II regions or spiral-arm structure. Conversely, the indirect estimates depend on assumptions about cluster membership, gamma-ray beaming, or empirical luminosity correlations.

\begin{figure}[htb]
\centering
\includegraphics[width=0.92\columnwidth]{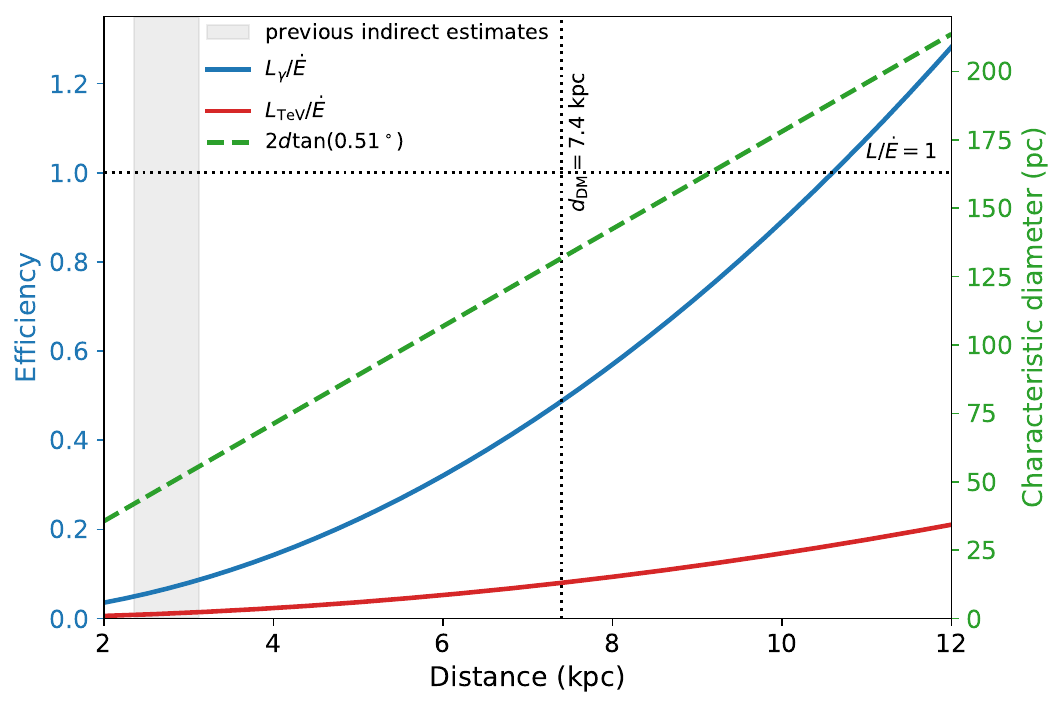}
\caption{Distance dependence of the inferred energetics and size of 1LHAASO J2238+5900. The solid curves show $L_\gamma/\dot{E}$ and $L_{\rm TeV}/\dot{E}$ for $f_\Omega=1$ and the WCDA spectral integration used here; the dashed curve shows the characteristic diameter $2d\tan(0.51^\circ)$(the KM2A scale is $2d\tan(0.43^\circ)$ $\simeq$ 111pc). The shaded band marks the range of previous indirect estimates (2.35–3.12 kpc), while the vertical dotted line marks the adopted DM-based distance. The high-distance tail of the DM estimate would imply a very high GeV efficiency, illustrating why a geometric parallax or phase-connected timing solution remains important.}
\label{fig:distance_implications}
\end{figure}

The GeV efficiency provides a useful physical consistency check. As shown in Figure~\ref{fig:distance_implications}, for $f_\Omega=1$, the fiducial DM distance implies $L_\gamma/\dot{E}\simeq0.49$, while the high-distance tail would approach or exceed unity. This argues that the far tail of the DM-distance estimate is physically disfavored unless the beaming correction is substantially below unity. The figure also illustrates how the inferred TeV efficiency and characteristic diameter vary with distance, and how the range of previous indirect estimates ($\sim$2.3--3.1 kpc, shaded band) would yield substantially smaller physical scales. Future VLBI astrometry, a phase-connected timing solution, or an independent X-ray absorption constraint will be needed to determine the true distance and calibrate electron-density models along this sightline.

\subsection{A Remark on the Youthfulness of PSR J2238+5903}

With $\tau_{\rm char}=26.6$ kyr, PSR J2238+5903 is young compared to the canonical TeV halo pulsars Geminga ($\tau_{\rm char}=342$ kyr) and Monogem ($\tau_{\rm char}=111$ kyr) \citep{Abeysekara2017}, and comparable in age to PSR B1823--13 (21 kyr), which powers the giant gamma-ray nebula HESS J1825--137 \citep{HESS2018}. If 1LHAASO J2238+5900 is a TeV halo, it would be one of the youngest such systems known, suggesting that particle escape from the PWN can occur at early evolutionary stages—possibly facilitated by a low-density ambient medium or early interaction with the reverse shock of the supernova remnant \citep{Chevalier2011}. The non-detection of a supernova remnant associated with this young pulsar, noted by \citet{Dincel2024}, is consistent with this picture: a rapidly expanding SNR in a tenuous medium could have already merged with the ISM or become undetectable.

The large extent and high TeV efficiency implied by our revised distance, together with the pulsar's young age, strengthen the interpretation of 1LHAASO J2238+5900 as a relic PWN in transition to a TeV halo.

\section{Conclusions}

We have detected $\mu$Jy-level radio pulsations from the gamma-ray pulsar PSR J2238+5903 using a 3000 s FAST L-band observation. The signal is recovered independently by FFT-based and FFA-based searches, peaks at $\mathrm{DM}=247.5\pm3.0$ pc cm$^{-3}$, and is absent from the simultaneous off-source beams. The inferred phase-averaged flux density, $S_{1250}\simeq3\,\mu$Jy, makes the pulsar an extremely radio-faint member of the Fermi pulsar population.

The measured DM provides the first pulsar-tied distance constraint for the PSR J2238+5903 / 1LHAASO J2238+5900 system. Standard Galactic electron-density models give distances of 6.3--9.0 kpc; adopting 7.4 kpc as a fiducial value, the LHAASO WCDA 39\% containment scale corresponds to a characteristic diameter of $\sim132$ pc and the $>1$ TeV luminosity is $\sim7.1\times10^{34}$ erg s$^{-1}$, about 8\% of the pulsar spin-down power.

A simple diffusion estimate using the KM2A extension gives $D\simeq8.5\times10^{28}$ cm$^2$ s$^{-1}$ at $\sim160$ TeV, comparable to the suppressed diffusion coefficients inferred for other TeV halos and well below the Galactic average.

These results make 1LHAASO J2238+5900 a promising young relic-PWN / TeV-halo transition candidate and demonstrate FAST's ability to uncover extremely faint radio counterparts of LHAASO-associated gamma-ray pulsars. Future FAST timing and polarization measurements, deeper X-ray and LHAASO observations, and a geometric distance constraint will be essential for determining the true distance and particle-transport interpretation of the system.

\begin{acknowledgments}
This work was supported by the National SKA Program of China (Nos. 2025SKA0140100, 2025SKA0110103); the National Natural Science Foundation of China (NSFC) grant Nos. 12375108, U1931111. This work made use of data from FAST (\url{https://cstr.cn/31116.02.FAST}). FAST is a Chinese national mega-science facility, operated by the National Astronomical Observatories, Chinese Academy of Sciences. We would like to thank the LHAASO Observatory (CSTR: \href{https://cstr.cn/31117.02.LHAASO}{31117.02.LHAASO}) and the LHAASO collaboration for valuable discussions.
\end{acknowledgments}

\bibliography{sample701}{}
\bibliographystyle{aasjournalv7}

\end{document}